\documentclass[aip, amsmath,amssymb,reprint]{revtex4-1}

\usepackage{dcolumn}
\usepackage{bm}
\usepackage[utf8]{inputenc}
\usepackage[T1]{fontenc}
\usepackage{mathptmx}
\usepackage{mathrsfs}
\usepackage{newsymbol}
\let\mathcal \undefined
\def\mathcal{\mathscr}

\let\ge       \undefined
\let\le       \undefined
\newsymbol\le          1336  
\newsymbol\ge          133E  
\newsymbol\notle       230A
\newsymbol\notge       230B

\newtheorem{theorem}{Theorem}
\newtheorem{remark}[theorem]{Remark}

\newtheorem{lemma}[theorem]{Lemma}
\newtheorem{proposition}[theorem]{Proposition}

\def\N{{\mathbb N}}
\def\Z{{\mathbb Z}}

\def\R{{\mathbb R}}

\newcommand{\Dom}{\mathsf{D}}
\newcommand{\la}{\lambda}
\newcommand{\ud}{\,{\rm d}}

\newcommand{\n}{\Vert}
\newcommand{\one}{{{\bf 1}}}

\newcommand{\ov}{\overline}
\newcommand{\iprod}[2]{( #1|#2 )}

\begin{document}

\title[The Garrison--Wong quantum phase operator revisited]{The Garrison--Wong quantum phase operator revisited}

\author{Jan van Neerven}
 \affiliation{Delft University of Technology, Faculty EEMCS/DIAM, P.O. Box 5031, 2600 GA Delft, The Netherlands}%
 \email{J.M.A.M.vanNeerven@TUDelft.nl}

\date{\today}

\begin{abstract}
We revisit the quantum phase operator $\Phi$ introduced by Garrison and Wong.
Denoting by $N$ the number operator, we provide a detailed proof of the Heisenberg commutation relation $\Phi N - N\Phi   = iI$
on the natural maximal domain $\Dom(\Phi N) \cap \Dom(N\Phi)$ as well as the failure of the Weyl commutation relations, and discuss some further interesting properties of this pair.
\end{abstract}

\maketitle

\section{Introduction}
A {\em Heisenberg pair} is an ordered pair $(A,B)$ of (possibly unbounded) self-adjoint operators, acting on the same Hilbert space $H$, such that for all $h\in \Dom(AB)\cap\Dom(BA)$ we have
$$ AB h - BA h = ih.$$
Here, $\Dom(AB) = \{h\in \Dom(B):\, Bh\in \Dom(A)\}$ and similarly the other way around. A {\em Weyl pair} is an ordered pair $(A,B)$ of (possibly unbounded) self-adjoint operators, acting on the same Hilbert space $H$, such that for all $t,s\in \R$ we have the operator identity
$$ e^{isA}e^{itB} = e^{-ist} e^{itB}e^{isA}.$$
Here, $(e^{isA})_{s\in\R}$ and $(e^{itB})_{t\in\R}$ are the strongly continuous one-parameter groups generated by $iA$ and $iB$ in the sense of Stone's theorem.
By straightforward differentiation (see Ref. \cite{Kato}) every Weyl pair is seen to be a Heisenberg pair, but the converse is false unless $A^2+B^2$ is essentially self-adjoint (this is the Rellich--Dixmier theorem, see Ref. \cite[Theorem 4.6.1]{Putnam} for a precise statement).

The standard textbook example (see Ref. \cite[Section 12.2]{Hall},
Ref. \cite[Section 2.11]{Putnam}) of a Heisenberg pair that is not a Weyl pair is the pair $(A,B)$ on $H= L^2(\mathbb{T})$ (where $\mathbb{T}$ is the unit circle in the complex plane) given by $$Af(\theta) = \theta f(\theta), \quad B f(\theta) = \frac1i f'(\theta).$$
This is a variation of the standard position-momentum pair.

The aim of this short note is to revisit, from a mathematician's point of view, some well known facts about the number-phase pair $(N,\Phi)$ on the Hilbert space $H^2(\mathbb{D})$ (where $\mathbb{D}$ is the open unit disc in the complex plane; the relevant definitions are given below). The interest of this pair derives from it being another example of a Heisenberg pair that is not a Weyl pair. This fact is well known and
contained in Ref. \cite{GarWon70}, except for some details concerning domains which we provide here. We also point out some further interesting features of this pair, providing along the way a rigorous justification of some observations in Ref. \cite{BGL}. As such this note does not contain new results, but we hope that it could be of some use to the more mathematically inclined reader interested in the subject.

\section{Quantum phase and Number}
The problem of defining quantum phase operators has been considered by many authors \cite{SusGlo64, GarWon70, BarPeg88, BarPeg89,  NFM92, BGL, LP2000, Busch2001, Busch2016} and has been reviewed in several places \cite{BarPeg92, Lyn95, PegBar97}. It has recently  found application in the context of quantum computing and quantum error correcting codes \cite{GKP, RKSE, GCB}. The proposal by Garrison and Wong \cite{GarWon70} is particularly attractive from a mathematical perspective. On the Hilbert space $$H:= H^2(\mathbb{D})$$ whose elements consist of the holomorphic functions $f(z) = \sum_{n\in\N} c_n z^n$ on the unit disc $\mathbb{D}$ for which
$$ \n f\n^2:= \sum_{n\in\N} |c_n|^2$$
is finite, they consider the bounded self-adjoint Toeplitz operator $\Phi$ with symbol $\arg(z)$, defined
for functions $f,g\in H$ by the relation
\begin{align}\label{eq:defPhi} \iprod{\Phi f}{g} = \frac1{2\pi} \int_{-\pi}^\pi \theta f(e^{i\theta}) \ov{g(e^{i\theta})}\ud \theta.
\end{align}
Here we identify the function $f(z)=\sum_{n\in\N} c_n z^n$ with the Fourier series $f(e^{i\theta}) = \sum_{n\in\N} c_n e^{in\theta}$ and similarly for $g$.
The choice of $\Phi$ as the quantum phase operator has been critically evaluated on physical grounds by several authors \cite{BerEng91, BarPeg92, GMT92, BGL}.

The {\em number operator} is the unbounded self-adjoint operator $N$ in $H$ given by
$$ Nf(z) = z f'(z)$$ on its maximal domain
$$\Dom(N) = \Bigl\{f = \sum_{n\in \N} c_n e_n \in H:\, \sum_{n\in\N} n^2|c_n|^2 <\infty\Bigr\}.$$
The spectrum of $N$ is given by $\sigma(N) = \N := \{0,1,2,\dots\}$ and
$$Ne_n = ne_n, \quad n\in\N,$$
where the functions $$e_n(z):= z^n, \quad n\in\N,$$ form an orthonormal basis of eigenvectors in $H$.

In what follows we write $[\Phi,N] := \Phi N  - N\Phi$ for the commutator of $\Phi$ and $N$, which we view as an operator defined on its  {\em maximal domain}
\begin{align*} \Dom([\Phi,N]) &:= \Dom(\Phi N) \cap \Dom(N\Phi) =  \Dom(N) \cap \Dom(N\Phi),
\intertext{where}
\Dom(\Phi N) & := \{f\in \Dom(N):\, Nf\in \Dom(\Phi) = H\} = \Dom(N),\\
\Dom(N\Phi) & := \{f\in \Dom(\Phi) = H: \, \Phi f \in \Dom(N)\}.
\end{align*}

\section{Main result}
It was shown by Garrison and Wong \cite{GarWon70} that the Heisenberg commutation relation
$$ [\Phi,N]f := (\Phi N  - N\Phi) f = if$$
holds for all functions $f$ in a suitable subspace $Y$, introduced in Lemma \ref{lem:dense}, which is dense in $H$ and contained in the domain of the commutator $[\Phi,N]$. This fact, which we take for granted for the moment, self-improves as follows.

\begin{proposition}\label{prop:Heisenberg-main1}
For all
$f\in \Dom([\Phi,N])$ one has $ [\Phi,N]f = if.$
\end{proposition}

\noindent{\em Proof.}
Let us denote by $A$ and $B$ the operator $[\Phi,N]$ with domains
$\Dom(A) = Y$ and $\Dom(B) = \Dom([\Phi,N])$. Then both $A$ and $B$ are densely defined and we have $A\subseteq B$. By Lemma \ref{lem:Heisenberg-necsuff}, $A$ is simply the restriction of the bounded operator $iI$ to $Y$. This operator is closable and since $Y$ is dense its closure equals $\ov A = iI$ with domain $\Dom(\ov A) = H$.

The self-adjointness of $\Phi$ and $N$ immediately implies that
\begin{align*}\iprod{i[\Phi,N]f}{g} = i(\iprod{Nf}{\Phi g} - \iprod{\Phi f}{N g}) = \iprod{f}{i[\Phi,N]g}
\end{align*}for all $f,g\in \Dom([\Phi,N])$. This means that $iB$ is symmetric. In particular, $iB$ (and hence $B$) is closable, a closed extension being given by its adjoints.

It now follows that $iI = \ov A \subseteq \ov B$ and therefore we must have $\Dom(\ov B) = H$. As a result, $\ov B = \ov A = iI$, and the asserted result follows.
\hfill$\square$\medskip

In the terminology introduces earlier, the proposition says that $(\Phi,N)$ is a Heisenberg pair. That it is not a Weyl pair can be seen by checking against the conditions of the Rellich--Dixmier theorem (as in Ref. \cite{GarWon70}) or by noting that the Stone--von Neumann uniqueness theorem (see Ref. \cite[Chapter 14]{Hall}) implies that both operators in a Weyl pair must be unbounded.

\begin{remark} {\rm We could generalise the definition of a Heisenberg pair by insisting only that the commutation relation $ AB h - BA h = ih$ hold for all $h\in Y$, where
$Y$ is some given dense subspace of $H$ contained in $\Dom(AB)\cap\Dom(BA)$. The above proof can be repeated {\em verbatim} to show that this definition, which is the one used in Ref. \cite{GarWon70},
is equivalent to the one given in the Introduction.}
\end{remark}

\begin{remark}{\rm
The following observation serves to justify our approach of interpreting the commutator $[\Phi,N]$ in terms of its maximal domain:
{\em Let $iN$ be the generator of a bounded $C_0$-group on a Banach space $X$. There does not exist a bounded linear operator $T$ on $X$ with the following two properties:
\begin{enumerate}
 \item[\rm(i)] for all $x\in \Dom(N)$ one has $Tx \in \Dom(N)$;
 \item[\rm(ii)] the identity $TNx-NTx= ix$ holds for all $x\in \Dom(N)$.
\end{enumerate}
}
\noindent Indeed, this is an immediate consequence of the second part of Ref. \cite[Theorem 3]{Vu} to $A = B = iN$ and $C = I$.
In our setting where $N$ is the number operator, the arguments in the preceding remark imply that the operator $\Phi$ fails property (i) for the function $x = \one$, the constant-one function.
}
\end{remark}

Let us now give a detailed derivation of the Garrison--Wong result, filling in some domain issues along the way. We split the result into two lemmas, Lemmas \ref{lem:Heisenberg-necsuff} and \ref{lem:dense}.
The starting point is the following explicit representation for $\Phi$, which follows readily from \eqref{eq:defPhi}:
$$ \iprod{\Phi e_m}{e_n} = \frac1{2\pi} \int_{-\pi}^\pi \theta e^{i(m-n)\theta}\ud\theta
= -i\frac{(-1)^{m-n}}{m-n}\delta_{m\not=n}.$$
Since $Ne_n = ne_n$, this gives
\begin{align*}
\ & \iprod{Ne_m}{\Phi e_n} - \iprod{\Phi e_m}{Ne_n}
 \\ & \qquad = -i\Bigl(\frac{ (-1)^{m-n}m}{m-n}- \frac{(-1)^{m-n}n}{m-n} \Bigr) \delta_{m\not=n}
 \\ & \qquad = -i(-1)^{m-n} \delta_{m\not=n}.
\end{align*}
It follows that if $f,g\in \Dom(N)$ are finite sums of the form $f = \sum_{j=0}^\ell c_j e_j$ and $g = \sum_{j=0}^\ell d_j e_j$, then
\begin{align*}
\ &  \iprod{Nf}{\Phi g}-  \iprod{\Phi f}{Ng}
\\ & \qquad = -i\sum_{j,k=0}^\ell (-1)^{j-k} c_j\ov {d_k} \delta_{j\not=k}
\\ & \qquad = i\iprod{f}{g} - i\sum_{j=0}^\ell c_j\ov{d_j} -i\sum_{j,k=0}^\ell (-1)^{j+k}c_j\ov {d_k}  \delta_{j\not=k}
\\ & \qquad = i\iprod{f}{g} - i\bigl(\sum_{j=0}^\ell (-1)^j c_j\Bigr)\ov{\Bigl(\sum_{k=0}^\ell (-1)^k d_k\Bigr)}.
\end{align*}
For arbitrary $f = \sum_{j\in\N} c_j e_j$ and $g = \sum_{j\in \N} d_j e_j$ in $\Dom(N)$ (with convergence of the sums in $H$)
we consider the truncations $f_\ell = \sum_{j=0}^\ell c_j e_j$ and $g_\ell = \sum_{j=0}^\ell d_j e_j$,
which satisfy $f_\ell,g_\ell\in \Dom(N)$ and $f_\ell\to f$ and $g_\ell \to g$ in the graph norm of $\Dom(N)$. In combination with the boundedness of $\Phi$ this gives
\begin{align*}
\ &  \iprod{Nf}{\Phi g} - \iprod{\Phi f}{Ng}
\\ & \qquad = \lim_{\ell\to\infty} (\iprod{N_\ell f}{\Phi g_\ell} - \iprod{\Phi f_\ell}{Ng_\ell})
\\ & \qquad =  i\iprod{f}{g} - i \Bigl(\sum_{j\in\N} (-1)^j c_j\Bigr)\ov{\Bigl(\sum_{k\in \N} (-1)^k d_k\Bigr)},
\end{align*}
where the limits in the last step exist by the absolute summability
$$\sum_{j\in\N} |c_j| \le \Bigl( \sum_{j\in\N} \frac1{(j+1)^2}\Big)^{1/2}
\Bigl( \sum_{j\in\N} {(j+1)^2}|c_j|^2\Big)^{1/2}
$$ using the Cauchy--Schwarz inequality. Both terms in the right-hand side product are finite,  the second because we are assuming that $f\in \Dom(N)$. In particular, if $f = \sum_{j\in \N}c_j e_j$ belongs to $ \Dom(N)$, then the series defining $f$ converges absolutely on $\ov{\mathbb{D}}$, and therefore such functions extend continuously to $\ov{\mathbb{D}}$.

The following lemma gives a necessary and sufficient condition for functions $f\in \Dom(N)$ to
satisfy the Heisenberg commutation relation. It provides some details for Ref. \cite[Eq. (4.8)]{GarWon70} as well as a converse to it.

\begin{lemma}\label{lem:Heisenberg-necsuff} For a function $f = \sum_{j\in\N} c_j e_j$ in $\Dom(N)$
the following assertions are equivalent:
\begin{enumerate}
\item[\rm(1)] $f\in \Dom([\Phi,N])$ and $[\Phi,N]f = if$;
\item[\rm(2)] $\displaystyle\sum_{j\in\N} (-1)^j c_j = 0$;
\item[\rm(3)] $f(-1)=0$.
\end{enumerate}
\end{lemma}

\noindent{\em Proof.}
The equivalence (2)$\Leftrightarrow $(3) is clear by the preceding observations.

\smallskip
(1)$\Rightarrow $(2): \
In the converse direction, if $f\in \Dom(N)$ belongs to $\Dom([\Phi,N])$ and $[\Phi,N]f = if$, then the above computation gives
\[ i\iprod{f}{f} = \iprod{[\Phi,N]f}{f} = i\iprod{f}{f} - i \Bigl|\sum_{j\in\N} (-1)^j c_j\Bigr|^2\]
and therefore $\sum_{j\in\N} (-1)^j c_j = 0.$

\smallskip
(2)$\Rightarrow $(1): \
Let $f,g\in \Dom(N)$ and suppose that $f = \sum_{j\in\N} c_j e_j$ with $\sum_{j\in\N} (-1)^j c_j = 0.$
 The above computation then gives
 $$ \iprod{Nf}{\Phi g}  - \iprod{\Phi f}{Ng} =  i\iprod{f}{g} $$
and therefore
$$ |\iprod{\Phi f}{Ng}| \le  (\n \Phi Nf\n +\n f\n)\n g\n. $$
This bound shows that $\Phi f\in \Dom(N^\star) = \Dom(N)$, which subsequently gives
$f\in \Dom([\Phi,N])$ and
$$  \iprod{[\Phi,N]f}{g} = \iprod{\Phi Nf}{g} - \iprod{N\Phi f}{g} =  i\iprod{f}{g}. $$
This being true for all $g$ in the dense subspace $\Dom(N)$, it follows
that $[\Phi,N]f = if$.
\hfill$\square$\medskip

These results imply the following curious cancellation result:
{\em
If a sequence of complex scalars $(c_n)_{n\in \N}$ satisfies
\begin{enumerate}
    \item[\rm(i)] $\displaystyle\sum_{n\in \N} n^2 |c_n|^2 < \infty$;
    \item[\rm(ii)] $\displaystyle \sum_{n\in \N} n^2\Big|\sum_{\substack{m\in \N \\ m\not=n}}\frac{(-1)^{m-n}}{m-n}c_m\Big|^2 < \infty$,
\end{enumerate}
then $$\sum_{n\in \N} (-1)^n c_n =0.$$
}
To see this, note that by \eqref{eq:Phi-em}, for functions $f = \sum_{n\in \N} c_n e_n$ we have
\begin{align*}
\Phi f (z) & = \sum_{m\in \N} c_m \Phi e_m(z)
 = -i \sum_{n\in \N}\Bigl(\sum_{\substack{m\in \N \\ n\not=m}}\frac{(-1)^{m-n}}{m-n}c_n \Bigr)z^n
\end{align*}
after changing the order of summation.
Thus (i) and (ii) say that $f\in \Dom(N)$ and $f\in \Dom(N\Phi)$, respectively, so together they say that $f\in \Dom([\Phi,N])$. For such functions,
Proposition \ref{prop:Heisenberg-main1} asserts that the Heisenberg commutation relation holds, and therefore the stated conclusion holds by virtue of Lemma \ref{lem:Heisenberg-necsuff}.

The next lemma from Ref. \cite{GarWon70} implies that $\Dom([\Phi,N])$ is dense in $H$.
The simple proof is included for the sake of completeness.

\begin{lemma}\label{lem:dense}
The subspace $Y$ of $H$ consisting of all functions
$f \in \Dom(N)$ satisfying the equivalent conditions
of Lemma \ref{lem:Heisenberg-necsuff} is dense in $H$.
\end{lemma}
\noindent{\em Proof.}
By the lemma \ref{lem:Heisenberg-necsuff}, for all integers $k\ge 1$
the function $f_k:= e_0 + \sum_{j=0}^{k-1} \frac1k e_{2j+1}$ belongs to $Y$.
Moreover we have $\n e_0 - f_k\n^2 = 1/k$. As a result, $e_0$ belongs to the closure $\overline{Y}$ of $Y$ in $H$.
Again by the lemma, for all $n\in \N$
we have $e_n+e_{n+1}\in Y$. This implies that $e_n\in \overline{Y}$ for all $n\in \N$, and therefore $Y$ is dense in $H$.
\hfill$\square$\medskip

It follows from Lemma \ref{lem:dense}, $\Dom([\Phi,N])$ is dense in $H$. In the light of this, the following negative result is perhaps somewhat surprising.

\begin{proposition} The domain $\Dom([\Phi,N])$ is not dense in $\Dom(N)$ with respect to the graph norm of the latter.
\end{proposition}
\noindent{\em Proof.}
We begin by observing that
\begin{align}\label{eq:Phi-em}\Phi e_m(z) = -i\sum_{\substack{n\in \N \\ n\not=m}}\frac{(-1)^{m-n}}{m-n}z^n.
\end{align}
By taking $m=0$, for $e_0 = {\bf 1}$ this gives
\begin{align}\label{eq:Phi-e0} \Phi{\bf 1}(z) = i\sum_{n=1}^\infty\frac{(-1)^{n}}{n}z^n = -i\log (1+z).
\end{align}

For all $g\in \Dom([\Phi,N])$ we have, using that $N^\star \one = N\one =0$,
$$ \iprod{\one}{[\Phi ,N]g}  = \iprod{\one}{\Phi Ng}   = \iprod{\Phi \one}{Ng}.$$
By the definition of adjoint operators, we have $\one \in \Dom([\Phi,N]^\star)$ if and only if there exists a constant $C$ such that for all $g\in \Dom[\Phi,N])$ we can estimate $|\iprod{\one}{[\Phi ,N]g}| \le C||g||$. If that is the case, we also obtain
that
\begin{align}\label{eq:rem}
|\iprod{\Phi \one}{Ng}|\le C||g||, \qquad g\in \Dom[\Phi,N]).
\end{align}

Suppose now, for a contradiction, that $\Dom([\Phi ,N])$ is dense in $\Dom(N)$ with respect to the graph norm. Then, by density, \eqref{eq:rem} implies the stronger statement
\begin{align*}
|\iprod{\Phi \one}{Ng}|\le C||g||, \qquad g\in \Dom(N).
\end{align*}
But this is equivalent to asserting that $\Phi \one \in \Dom(N^\star) = \Dom(N)$.
But in that case $\Phi \one(z) = -i\log(1+z)$ (cf. \eqref{eq:Phi-e0}) would extend continuously
to $\ov{\mathbb{D}}$, which is not the case.

Since $\Dom([\Phi,N])$ is dense
in $H$ the adjoint operator $[\Phi,N]^\star$ is well defined, and
the preceding argument proves that if $\Dom([\Phi ,N])$ is dense in $\Dom(N)$, then
$\one\not\in \Dom([\Phi,N]^\star)$.
But then we arrive at the contradiction
$$ \one \not \in  \Dom([\Phi,N]^\star) =  \Dom(\ov{[\Phi,N]}^\star) = \Dom((iI)^\star) = H,$$
using Ref. \cite[Theorem 1.8]{Schm} to justify the first equality.
\hfill$\square$\medskip

Let $V$ be the contraction on $H$ defined by the left shift
$$ Ve_0 := 0, \quad Ve_n := e_{n-1}, \quad  n\ge 1.$$
Identifying the functions $e_n$ with elements of $L^2(\mathbb{T})$,
the two-sided left shift on $L^2(\mathbb{T})$ is a unitary extension of $V$ and is therefore
given by a unique projection-valued measure $P$ on $\mathbb{T}$. Compressing $P$ to
$H$ produces a positive operator-valued measure (POVM) $Q$ on $\mathbb{T}$
such that
$$ V^k = \int_{\mathbb T} \la^k\ud Q(\la), \quad k=0,1,2\dots,$$
and this property uniquely characterises $Q$ as a POVM. For the details the reader is referred to Ref. \cite{AkhGla2}; see also \cite{Berberian, Holevo}.

As implicitly observed on page 87 of Ref. \cite{BGL}, the Garrison--Wong operator $\Phi$ can be characterised in terms of $Q$ as follows.

\begin{proposition}\label{prop:GW-repr}
$\displaystyle\Phi =\int_{\mathbb{T}} \arg(\la)\ud Q(\la).$
\end{proposition}
The rigorous interpretation of the integral on the right-hand side is as follows. Denoting by $B_{\rm b}(\mathbb{T})$
the Banach space of all bounded Borel measurable functions on $\mathbb{T}$, one uses the boundedness of the Borel calculus of the projection-valued measure $P$ associated with $V$ to obtain that there exists a unique linear mapping $\Psi:B_{\rm b}(\mathbb{T})\to \mathscr{L}(H)$, the space of bounded operators on $H$,
satisfying
 $$ \Psi(\one_B) = Q_B, \quad B\subseteq \mathbb{T} \ \hbox{ Borel},$$
  and
 $$ \n \Psi(f)\n \le \n f\n_\infty, \quad f\in B_{\rm b}(\mathbb{T}).$$
 It further satisfies $$\Psi(f)^\star = \Psi(\ov f), \quad f\in B_{\rm b}(\mathbb{T}).$$
We now define
$$ \int_{\mathbb{T}} f\ud Q := \Psi(f), \qquad f\in  B_{\rm b}(\mathbb{T}).$$
For all $f_1,f_2\in A(\mathbb{T})$, the uniform closure of the trigonometric polynomials in $C(\mathbb{T})$, we have
 $$  \Psi(f_1) \Psi(f_2)=  \Psi(f_1f_2),$$
but this property does not extend to general functions $f_1,f_2\in B_{\rm b}(\mathbb{T})$.

\medskip\noindent{\em Proof of Proposition \ref{prop:GW-repr}.}
We equipartition $\mathbb{T}=(-\pi,\pi]$ into $k$ subintervals of length $2\pi/k$ by setting
$I_j := (-\pi+2\pi(j-1)/k, -\pi + 2\pi j/k]$ for $j=1,\dots,k$. Then, by the continuity of  $\Psi: f\mapsto \int_{\mathbb{T}} f\ud Q$,
 \begin{align*}
\int_{\mathbb{T}} \arg(\la)\ud Q(\la)
  & = \lim_{k\to\infty} \int_{\mathbb{T}} \arg_k(\la)\ud Q(\la)
  \\ & = \lim_{k\to\infty} \sum_{j=1}^k \frac{2\pi j}{k} Q(I_j),
 \end{align*}
 where $ \arg_k(\la) := \sum_{j=1}^N \frac{2\pi j}{k}{\bf 1}_{I_j}(\la)$.
To compute $Q(I_j)$ we use that $Q$ is the compression to $H$ of the projection-valued measure $P$ associated
with the two-sided shift $U$ on $L^2(\mathbb{T})$. The latter is given by $P(I_n)f = \one_{I_n}f$ for $f\in L^2(\mathbb{T})$.
Accordingly, if we denote the inclusion mapping $H\mapsto L^2(\mathbb{T})$ by $J$, then for all $f\in H$ we have
$$ Q(I_j)f = J^\star P(I_j)Jf = J^\star \one_{I_j}f.$$
It follows that
$$ \int_{\mathbb{T}} \arg(\la)\ud Q(\la) = \lim_{k\to\infty} \sum_{j=1}^k \frac{2\pi j}{k} J^\star {\bf 1}_{I_j},$$
identifying ${\bf 1}_{I_j}$ with the multiplication operator $f\mapsto {\bf 1}_{I_j}f$ from $H$ to $L^2(\mathbb{T})$.
On the other hand, by the definition of the operator $\Phi$,
$$ (\Phi f|g) = \int_{\mathbb{T}}\arg(\la) f(\la) \ov{g(\la)} \ud \la, \quad f,g\in H,$$
we have
\begin{align*}
\Phi f  & =  J^\star (\arg(\cdot) f(\cdot))
 \\ & =  \lim_{k\to\infty} J^\star (\arg_k(\cdot) f(\cdot))
 = \lim_{k\to\infty} \sum_{j=1}^k \frac{2\pi j}{k} J^\star ({\bf 1}_{I_j}f).
\end{align*}
This completes the proof.
\hfill$\square$\medskip

\begin{remark}
{\rm The arguments used in the proof imply that $Q_B = 0$ if the Borel set $B$ has measure $0$. It follows that
the integral $\int_{\mathbb{T}} f\ud Q$ is well defined for functions $f\in L^\infty(\mathbb{T})$. With essentially the same proof as above one shows that for any $\phi\in L^\infty(\mathbb{T})$ the
bounded Toeplitz operator $T_\phi$ on $H$ with symbol $\phi$ is given by
$$ T_\phi = \int_{\mathbb{T}} f \ud Q.$$
}
\end{remark}

As observed in Ref. \cite{BGL},
the POVM $Q$ obeys the following ``covariance property''. For the reader's convenience we include the simple proof.

\begin{proposition} For all $t\in\R$ and Borel sets $B\subseteq \mathbb{T}$ we have
\begin{align*}
e^{itN}Q(B) e^{-itN} = Q({e^{it}B}),
\end{align*}
where $(e^{itN})_{t\ge 0}$ is the unitary $C_0$-group on $H$ generated by $iN$
and $e^{it}B = \{e^{it}\la: \, \la\in B\}$ is the rotation of $B$ over angle $t$.
\end{proposition}
\noindent{\em Proof.} The properties of the projection-valued measure $P$ used in the proof of Proposition \ref{prop:GW-repr} imply that for the trigonometric functions $e_k$, $k\in\N$, we have
\begin{align*}
 (Q(B) e^{-itN}e_n|e_m)  = (P(B) J e^{-itN}e_n|Je_m)=  e^{-int}({\bf 1}_B e_n|e_m)
\end{align*}
while at the same time
\begin{align*}
& (e^{-itN} Q({e^{it}B})e_n|e_m)
 \\ & \quad = (P(e^{it}B)Je_n|Je^{itN}e_m) =  e^{-itm}({\bf 1}_{e^{it}B}e_n|e_m)
 \\ & \quad = e^{-itm}\int_{e^{it}B}\la^{n-m}\ud\la = e^{-itm}\int_{B}(e^{-it}\mu)^{n-m}\ud\mu
 \\ & \quad = e^{-itn}\int_{e^{it}B}\mu^{n-m}\ud\mu = e^{-int}({\bf 1}_B e_n|e_m).
\end{align*}
Since the span of the trigonometric functions is dense in $H$, this completes the proof.
\hfill$\square$\medskip

This contrasts with the failure of the {\em Weyl commutation relations}
\begin{align}\label{eq:Weyl} e^{itN}e^{is\Phi} e^{-itN} = e^{-ist}e^{is\Phi}, \qquad s,t\in\R.\end{align}
This failure is usually demonstrated by showing that \eqref{eq:Weyl} would imply the identity $ a = e^{-i\Phi}N^{1/2}$,
where $a$ is the annihilation operator associated with $N$ (so that $a^\star a = N$); this identity is subsequently shown to be impossible if at the same time $\Phi$ is to be self-adjoint (see Ref. \cite{CarNie, GarWon70, SusGlo64}).

Here, by elementary methods, we will give a direct proof of the more precise result that the Weyl relation \eqref{eq:Weyl} fails for every fixed $s\not=0$:

\begin{proposition} Let $T$ be an arbitrary bounded operator on $H$.
 If $s\in \R$ is such that for all $t\in \R$ one has
\begin{align}\label{eq:s}
e^{itN}e^{isT} e^{-itN} = e^{-ist}e^{isT},
\end{align}
then $s = 0$.
The same conclusion holds if we assume that $T$ is a (possibly unbounded) self-adjoint operator on $H$.
\end{proposition}
\noindent{\em Proof.}
Suppose that  $s\in \R$ is such that \eqref{eq:s} holds for all $t\in \R$. Choose $n,m\in \N$ so  that $(e^{isT}e_n|e_m)\not=0$.
Applying \eqref{eq:s} to $e_n$ and taking inner products with $e_m$, we obtain
$$ e^{it(m-n)} (e^{isT} e_n|e_m)  = e^{-ist}(e^{isT}e_n|e_m).$$
This can hold for all $t\in \R$ only if $e^{it(m-n)}= e^{-ist}$ for all $t\in \R$, forcing $s = n-m\in\Z$.

Suppose next that $s = k\in \Z$ is such that \eqref{eq:s} holds for all $t\in \R$. If $k\ge 1$, the above argument shows that we must have
$(e^{isT} e_n|e_m) = 0$ unless $n-m = k$, which implies that $e^{isT} e_{n}$ is a multiple of $e_{n-k}$ if $n\ge k$
and $e^{isT} e_n = 0$ if $0\le n \le k-1$. Given a fixed $n\in \N$, it follows that $e^{imsT}e_n = 0$ for all sufficiently large $m\in\N$. But this is impossible as it would lead to the contradiction $$e_n = e^{-imsT}e^{imsT}e_n = 0.$$
If $k\le -1$, similar reasoning gives that $e^{isT} e_{n}$ is a multiple of $e_{n-k} = e_{n+|k|}$ for all $n\in\N$, so
$e^{ikT}f(z) = z^{|k|}f(z)$ for all $f\in H$. With $f =  e^{-ikT}e_j$ this leads to the contradiction that,
for $0\le j\le |k|-1$, $$e_j =  e^{ikT} e^{-ikT}e_j \in \overline{\hbox{span}}\{e_i: \, i\ge |k|\}.$$
\hfill$\square$\medskip

These considerations support the case, made in Ref. \cite{BGL}, that the POVM should be considered the ``correct'' phase observable (in the generalised sense of POVM's).

\bibliographystyle{amsplain}
\bibliography{phase}

\end{document}